# Three-dimensional label-free imaging and quantification of lipid droplets in live hepatocytes

**Kyoohyun Kim**[1], **SeoEun Lee**[1,a], **Jonghee Yoon**[1,b], **JiHan Heo**[1,c], **Chulhee Choi**[2], **and YongKeun Park**[1,3,*]

[1]Department of Physics, Korea Advanced Institute of Science and Technology, Daejeon 34141, Republic of Korea.
[2]Department of Bio and Brain Engineering, Korea Advanced Institute of Science and Technology, Daejeon 34141, Republic of Korea.
[3]TOMOCUBE, Inc., Daejeon 34051, Republic of Korea
[a]Present address: Graduate School of Arts and Sciences, Columbia University, NY 10032, USA
[b]Present address: Department of Physics, University of Cambridge, Cambridge CB3 0HE, UK
[c]Present address: College of Medicine, Seoul National University, Seoul 03080, Republic of Korea
    *Correspondence:
    Prof. YongKeun Park, E-mail: yk.park@kaist.ac.kr, Tel: +82-42-350-2514, Fax: +82-42-350-2510

**Lipid droplets (LDs) are subcellular organelles with important roles in lipid storage and metabolism and involved in various diseases including cancer, obesity, and diabetes. Conventional methods, however, have limited ability to provide quantitative information on individual LDs and have limited capability for three-dimensional (3-D) imaging of LDs in live cells especially for fast acquisition of 3-D dynamics. Here, we present an optical method based on 3-D quantitative phase imaging to measure the 3-D structural distribution and biochemical parameters (concentration and dry mass) of individual LDs in live cells without using exogenous labelling agents. The biochemical change of LDs under oleic acid treatment was quantitatively investigated, and 4-D tracking of the fast dynamics of LDs revealed the intracellular transport of LDs in live cells.**

Lipid droplets (LDs), a monolayer enclosing phospholipid and associated proteins, are main lipid-storing subcellular organelles in most cell types[1]. Recent studies suggest that LDs exhibit highly dynamic three-dimensional (3-D) motions inside cells to regulate intracellular lipid storage and provide a lipid source for energy metabolism[2-4]. In addition to their roles in lipid storage and metabolism, LDs are related with various pathologies, including cancer, obesity, and diabetes mellitus[5,6]. However, detailed mechanisms underlying the biosynthesis, growth, movement and interaction of LDs and their interactions with other organelles remain elusive. For these reasons, studying the 3-D intracellular motions of LDs under various pathophysiological conditions is important to identify the complex regulatory mechanisms of LDs.

To date, a number of techniques have been used to image intracellular LDs and their motions. Fluorescence microscopy can selectively visualize LDs by staining LDs with various fluorescent probes[7-11], but introducing the probes may perturb the physiological conditions of the LDs due to phototoxicity and photobleaching, and the chemicals used for the probe treatment and sample preparation can induce fusion of adjacent LDs[7]. Recently, coherent anti-Stokes Raman scattering (CARS) microscopy has been used for label-free imaging of LDs from the molecular vibrational signals of long $-CH_2$ lipid chains[12-15]. Additionally, CARS microscopy has provided tracking of LDs inside cells revealing their lipid transport mechanism[16,17]. However, CARS microscopy uses complicated and bulky optical components for nonlinear optical processes, and vibrational oscillators for target molecules with low concentrations have a low signal-to-noise ratio because the CARS signal is proportional to the square of the oscillator concentration. Moreover, both fluorescence microscopy and CARS microscopy require axial stacking for 3-D LD imaging, which limits the real-time investigation of fast 3-D dynamics of LDs inside cells.

Recently, quantitative phase imaging (QPI) techniques have emerged that provide quantitative morphological and biochemical information on individual cells and tissues without using exogenous labelling agents. QPI techniques use interferometry to measure complex optical fields consisting of the amplitude and phase delay of light diffracted by biological samples[18,19]. Measured 2-D optical phase delay of individual cells corresponds to the dry mass of each cell[20-22], which is the mass of the aqueous components inside the cell including proteins and subcellular organelles, and the temporal fluctuation of the optical phase delay is implemented to study the membrane fluctuation of biological cells on a nanometre level[23-25]. Moreover, the measurements of multiple complex optical fields with various incident angles reconstruct the 3-D refractive index (RI) distribution of biological samples, which corresponds to the intracellular protein concentration of individual cell[26-37]. The 3-D RI distribution of biological samples can serve as an alternate solution for 3-D label-free imaging and quantification of LDs in live cells. Because lipids have RI values distinctively higher than that of cytoplasm in cells[38], the 3-D RI distribution of LDs can be effectively segmented from surrounding cytoplasm, which also provides quantitative information about LDs without using labelling agents.

Here, we propose the measurement of the 3-D RI distribution of intracellular LDs in living cells for label-free quantitative imaging. A Mach-Zehnder interferometer measures complex optical fields of live hepatocytes (human hepatocellular carcinoma cell line, Huh-7) with various illumination angles, from which the 3-D RI distribution of the cells is reconstructed via optical diffraction tomography (ODT). Because ODT takes into account of light diffraction inside cells and intracellular LDs, highly scattering LDs with high RI values are successfully visualized in three dimensions. From the reconstructed 3-D RI distribution, structural and biochemical information is quantitatively measured including the concentration and the dry mass of lipid as well as the volume of the individual LDs under a chemical treatment. In addition, time-lapse measurements of the 3-D RI distribution of live cells are used to investigate the 4-D intracellular dynamics of individual LDs which reveal various complicated 3-D diffusive motions of the LDs inside cells.

Results

**Optical diffraction tomography for 3-D LD imaging**

In order to measure the 3-D RI distribution of LDs in live hepatocytes, ODT-based on Mach-Zehnder interferometry was exploited shown in Fig. 1a (see Methods). Multiple holograms of an individual cell with various illumination angles were measured by Mach-Zehnder interferometry (Fig. 1b), from which complex optical fields consisting of the amplitude (Fig. 1c) and phase delay (Fig. 1d) of the light diffracted by the sample were retrieved by a field retrieval algorithm based on Fourier transform[39,40]. Multiple complex optical fields from various illumination angles were used to reconstruct the 3-D RI distribution of the samples with the Fourier diffraction theorem[26,30,41] (see Methods). As shown in the cross-sectional slices of a reconstructed tomogram in Fig. 1e, LDs inside individual hepatocytes have significantly high RI values compared to the surrounding cytoplasm of cells. Moreover, isosurfaces of the reconstructed tomogram clearly reveal the 3-D spatial distribution of LDs inside individual hepatocytes (Fig. 1g and Supplementary Movie 1).

**Validation of ODT for LD imaging**

To confirm that the intracellular organelles with high RI values in the reconstructed tomograms represent LDs, we compared the 3-D RI distributions of 6 human hepatocyte cells with fluorescence images of the same cells stained with Nile Red (absorption at $\lambda = 488$ nm and emission at $\lambda = 570$ nm, 19123, Sigma-Aldrich Inc., USA). To measure the fluorescence images of LDs inside hepatocytes, samples were stained with 0.5 μg/ml of Nile Red for 15 min. in a humidified $CO_2$ incubator (5% $CO_2$/95% air at 37 ℃) and excited with a collimated light emitting diode (LED) (Nominal wavelength $\lambda = 455$ nm, 500 mW, M455L3-C1, Thorlabs Inc. USA) through a long-pass dichroic mirror (DMLP505R, Thorlabs Inc.) after the acquisition of complex optical fields from various illumination angles. The emitted fluorescence signals were collected with the same high NA objective lens and recorded with the same camera. Deconvolution of the fluorescence images was performed with a MATLAB built-in function, deconvblind, to enhance the image contrast from blurred images so that the positions of LDs could easily be identified.

The results show that subcellular organelles with RI values of $n = 1.375$ or higher correspond to the LDs labelled with fluorescent probes. By visual inspection, the positions of LDs stained with Nile Red in a deconvoluted fluorescence image in Fig. 1f are identical to the positions of regions with high RI values in the 3-D RI tomograms (Fig. 1e). For quantitative analysis, the centroid positions of 265 individual LDs in the deconvoluted fluorescence image of 6 hepatocytes were matched with the cross-sectional slice image of a 3-D RI tomogram in the focal plane, and 95.5% of the subcellular organelles with high RI values corresponded to LDs stained with the fluorescent probe. The result clearly shows that ODT has high sensitivity for the 3-D imaging of LDs inside individual hepatocytes.

**Effects of oleic acid on the morphological and physical properties of LDs**

The measured 3-D RI distribution of individual LDs inside hepatocytes provides quantitative biochemical properties of samples at high resolution. In order to demonstrate the capability of the present method for biochemical studies, we quantitatively measured the change in morphological and biochemical parameters of LDs under oleic acid (OA) treatment from the reconstructed 3-D RI distributions of the LDs inside the hepatocytes. OA is a long-chain free fatty acid and a major component of plasma and tissue lipid, which is known as a potent inducer of LD formation[42]. We calculated the dry mass ratio of LDs to the cell, the total volume ratio of LDs to the cell, and lipid concentration of LDs inside hepatocytes in response to the OA treatment from the reconstructed tomograms (see Methods).

The results with the OA treatment are shown in Fig. 2. By visual inspection, the 3-D isosurface and cross-sectional slices of the 3-D RI distributions of hepatocytes under OA treatment clearly show that the LDs are increased in number and volume after the OA treatment (Figs. 2b, 2d and Supplementary Movie 3) compared with LDs in untreated hepatocytes (Figs. 2a, 2c and Supplementary Movie 2). Quantitative analysis of the measured RI distribution shows that the OA treatment induces an increase in the total amount of LDs inside cells while the change in the lipid concentration of LDs, as well as the protein concentration in the cytoplasm of hepatocytes, is not significant. As shown in Fig. 2e, the mean RI value of the cytoplasm of the hepatocytes does not change under OA treatment from 1.347 ± 0.002 to 1.348 ± 0.002, which corresponds to a protein concentration of 4.99 ± 1.01 g/dL and 5.59 ± 1.16 g/dL, respectively. Additionally, the mean RI value of LDs is constant under the OA treatment from 1.386 ± 0.004 to 1.385 ± 0.004 which corresponds to a lipid concentration in the LDs of 36.5 ± 2.70 g/dL to 35.7 ± 2.68 g/dL, respectively (Fig. 2f). The results imply that the OA treatment does not affect the dry mass density of the cytoplasm and LDs.

In contrast, the volume ratio of LDs to cell increases from 0.011 ± 0.007 to 0.048 ± 0.025 after the OA treatment (Fig. 2g), and the dry mass ratio of LDs to cell increases from 0.062 ± 0.040 to 0.217 ± 0.090 after the OA treatment (Fig. 2h). Both results show that the volume and dry mass ratio of LDs to cell increase about four times by the OA treatment, and it is consistent with the previous observation[42]. Because the lipid concentration of the LDs is maintained after the OA treatment, the result implies that the most increase of the dry mass ratio of LDs is from an increase in the volume of the LDs. Quantitative analysis of individual LDs inside hepatocytes shows that the volume of individual LDs per cell increases from 0.34 ± 0.26 fL to 0.75 ± 0.30 fL (Fig. 2i), and the number of LDs per cell increases from 153.3 ± 86.13 to 270 ± 128.2 (Fig. 2j) after the OA treatment. The result indicates that the increase of the total volume of LDs after the OA treatment is resulted from the increase of both number and volume of individual LDs, and also suggests the potential application of the present method to the study of biogenesis of LD in individual cells[2,43].

**4-D RI distribution of the dynamics of LDs in live hepatocytes**

The present method can also be used to study the 3-D dynamics of LDs inside cells by measuring the time-lapse 3-D RI distribution of hepatocytes. We measured the 3-D RI tomograms of an individual hepatocyte every 3 seconds for 93 seconds. To enhance the tomogram acquisition speed, the galvanomirror circularly scanned 30 illumination beams with various azimuthal angles at a scanning rate of 0.3 seconds/cycle. The acquisition of 30 complex optical fields from various illumination angles is enough for a robust tomogram reconstruction; the tomograms reconstructed from 30 viewing angles, and 300 full viewing angles have a correlation coefficient of 0.95 (data not shown).

From the reconstructed time-lapse 3-D RI distribution of hepatocytes and LDs, 3-D particle tracking of individual LDs was performed. The 3-D positions of 51 individual LDs in five hepatocytes were tracked using a customized MATLAB script. The 3-D RI distributions at each time frame were first segmented by the RI threshold value ($n > 1.375$), and the 3-D centroid positions of individual LDs of interest were measured over time. As shown in Figs. 3a-b and Supplementary Movie 4, the time-lapse 3-D RI distributions of LDs inside a hepatocyte revealed various diffusive motions of LDs ranging from subdiffusion to active transport. For example, an LD in Fig. 3d diffused within the confined volume of 0.125 fL for 93 seconds shown by its 3-D trajectory (Fig. 3f), while another LD in Fig. 3e moved about 3 μm during the same time range (Fig. 3g).

In order to investigate the various diffusive motions of LDs inside the cell, we measured the mean-squared distribution (MSD) versus elapsed time relation from the 3-D trajectory of each LD, and the relation was fitted to the power-law equation[16,44,45], $\langle r^2(t) \rangle = 6Dt^\alpha$, where $D$ is the diffusion coefficient and $\alpha$ is the power-law exponent of anomalous diffusion. The results imply that the 3-D dynamic motion of LDs in Fig. 3f and Fig. 3g are subdiffusive and active transport motion, given by $\langle r^2(t) \rangle = 6Dt^{0.62}$ (Fig. 3h) and $\langle r^2(t) \rangle = 6Dt^{1.56}$ (Fig. 3i), respectively. The diffusion coefficients and the power-law exponents of the 3-D trajectories of 51 LDs are $D = 0.0014 ± 0.0016$ μm²/s and $\alpha = 0.99 ± 0.49$ (Fig. 3j), respectively. The broad range of the measured power exponents implies that LDs exhibit various anomalous diffusive motions from subdiffusion to active transport. The subcellular organelles may exhibit subdiffusive motion when thermal diffusion is restricted by a crowded cytoskeleton and extracellular matrix (ECM), while active transport motion is driven by molecular motor proteins traveling along the cytoskeleton[46-48]. The results clearly show that the present method reveals the complex 3-D dynamics of LDs inside live cells originating from the interaction between LDs and surrounding cellular architectures.

**Discussion**

In this report, we used QPI for label-free quantitative imaging of LDs in living cells. Interferometric microscopy measures optical phase delay induced by individual LDs inside living cells, and reconstructed 3-D RI distribution of LDs provides quantitative morphological and biochemical information (e.g. lipid concentration, volume, and dry mass) of individual LDs without isolating LDs from a cell. The present technique was used for the quantitative measurement of changes in the physiological parameters of LDs in response to OA treatment. In addition, time-lapse tomogram measurements revealed that the 3-D dynamics of LDs exhibited a variety of diffusion modes inside living cells, which are related to lipid transport and storage.

The present technique is a non-perturbative method that does not intrude on the physiological conditions of LDs and surrounding cellular components. Because optical phase delay and 3-D RI distribution of biological samples are intrinsic optical properties which are sensitive to morphological structures and chemical compositions, the present technique can measure the physical and chemical parameters of LDs quantitatively without the addition of exogenous fluorescent agents. We expect that the present method can be implemented to investigate lipid metabolism and related diseases including obesity, type II diabetes, hepatosteatosis, atherosclerosis, and Hepatitis C[49].

The high RI values of LDs enable the present technique to visualize individual LDs inside a human hepatocyte with high sensitivity when compared with fluorescence images of the same samples. The statistical analysis implies that around 95% of LDs stained by lipophilic fluorescent probes have high RI values in their tomograms. There is a 10.3% possibility that tomogram voxels with high RI values do not correspond to LDs. They can originate from other proteins and subcellular organelles with high RI values. In order to improve the molecular specificity, the measurements for RI dispersion of LDs using spectroscopic QPI can identify LDs from other proteins and subcellular structures with high RI values[50-52].

The sensitivity of RI measurements of our system was calculated as $\Delta n = 3.32 \times 10^{-4}$, by measuring the standard deviations of RI values in the background region (Supplementary Figure 1). This RI sensitivity implies that the present system can resolve the concentration of lipid in LDs as low as 0.175 g/dL. In comparison, the signal-to-noise ratio of CARS microscopy is determined by various parameters including signal integration time, input beam power, and the sensitivity of the detector. This sensitivity of the present method is comparable to or slightly better than a typical CARS microscope, which has the detection sensitivity of 8 mmol/L of the methanol concentration using the C–H stretch peak at 2,839 cm$^{-1}$, which corresponds to 0.64 g/dL of triglyceride (molecular weight = 799 g/mol)[53].

Given that an LD is a dynamic organelle but how it moves has been unknown, the present technique revealed the 3-D motions of individual LDs with the time-lapse measurements of 3-D RI distributions of LDs inside living cells taken every 3 seconds at a scanning rate of 0.3 sec/cycle. The 3-D trajectories of individual LDs exhibited various modes of diffusion ranging from subdiffusion intruded by the crowded cytoskeleton and cytoplasm to active transport powered by molecular motor proteins along the cytoskeleton. The present technique extends the dimension of LD tracking into three dimensions which have been mainly done in two dimensions with CARS imaging[19, 32], and the present method can be further improved for studying fast dynamics in long-term measurements of LDs. The current tomogram acquisition rate of 0.3 sec/cycle is fast enough to resolve the 3-D dynamics of individual LDs, and it can be further increased up to 16.7 milliseconds/cycle, which enables the measurements of fast dynamics of LDs[54]. In addition, the present technique enables the long-term 3-D tracking of LDs thanks to the label-free nature of the method without the addition of fluorescence agents, which may exhibit photobleaching and phototoxicity.

One on the limitations of the present method is that it requires for a setup with careful alignments for interferometry, and precise controls and measurements, compared to the existing fluorescence-based imaging that can be performed with relative ease. Nonetheless, recently 3D RI tomography system has been commercialized[55], which will further facilitate direct applications of this method to the fields.

In conclusion, we present a QPI technique providing label-free quantitative measurements of chemical and physiological parameters of LDs from the reconstructed 3-D RI distribution of living cells. The present method with the capability of 3-D LD tracking revealed the intracellular motion of LDs inside living cells without exogenous labelling agents, and we expect the present method can pave the way for revealing the biochemical mechanism of lipid storage and energy metabolism and related diseases.

**Methods**

**Sample preparation.** Human hepatocytes (human hepatocellular carcinoma cell line, Huh-7, Apath, Brooklyn, NY, USA) were prepared according to the standard protocol[56]. Briefly, Huh-7 cells were maintained in Dulbecco's modified Eagles' medium (DMEM, Gibco, Big Cabin, Oklahoma, USA) supplemented with 10% heat-inactivated fetal bovine serum, 4,500 mg/L D-glucose, L-glutamine, 110 mg/L sodium pyruvate, sodium bicarbonate, 100 U/ml penicillin, and 100 μg/mL streptomycin. Cells were subcultivated for 4 – 8 hours before the experiments on 24 × 40 mm cover glass (Marienfeld-Superior, 25814017), and then washed with Dulbecco's Phosphate Buffered Saline (DPBS, Gibco) before the measurements. Pre-warmed DPBS (400 μl) was added to the sample, and then another cover glass was put on the sample to prevent drying.

**Chemical treatment.** For a deconvoluted fluorescence imaging, intracellular lipid droplets were stained with 0.5 μg/ml of Nile Red (Sigma-Aldrich, 19123) for 15 min[57]. For the oleic acid treatment, cells were cultured in a medium containing 0.1% of filtered oleic acid (OA, Sigma-Aldrich, O1383) conjugated with fatty acid-free bovine serum albumin (BSA, Sigma-Aldrich). Procedures for BSA-OA formation were done according to a previous report[58]. Briefly, 1 ml BSA in DPBS (10 mg/ml) was mixed with 10 μl of OA in ethanol (100 mg/ml). The BSA-OA mixture was placed at room temperature for 30 min, and then filtrated with a 0.2 μm filter before addition to the culture medium.

**Optical setup.** In order to measure complex optical fields diffracted by a sample, QPI based on Mach-Zehnder interferometry was exploited (Fig. 1a). A diode-pumped solid-state laser ($\lambda$ = 532 nm, 50 mW, Samba$^{TM}$, Cobolt Co., Sweden) beam is divided into two arms by a beam splitter. One beam is used as a reference beam and the other beam illuminates samples on the stage of an inverted microscope (IX73, Olympus Inc., Japan) through a tube lens ($f$ = 200 mm) and a high numerical aperture (NA) condenser lens (NA = 0.9, UPLFLN 60 ×, Olympus Inc.). To reconstruct a 3-D RI tomogram of an individual cell, a cell is illuminated with 300 various incident angles scanned by a dual-axis galvanomirror (GVS012/M, Thorlabs Inc., USA). The diffracted beam from a sample is collected by a high NA objective lens (oil immersion, NA = 1.42, PLAPON 60 ×, Olympus Inc.) and a tube lens ($f$ = 180 mm). The beam is further magnified by an additional 4-$f$ lens system, and the total magnification is set to be 240×. The diffracted beam from the samples interferes with the reference beam at an image plane, and generates a spatially modulated hologram (Fig. 1b). The hologram is recorded by a high-speed scientific CMOS camera (Neo sCMOS, Andor Inc., UK) with a frame rate of 100 Hz.

**Optical diffraction tomography.** For the reconstruction of the 3-D RI distribution of biological samples, we implemented optical diffraction tomography based on the Fourier diffraction theorem[26,27,30,41]. According to the Fourier diffraction theorem, a 2-D Fourier spectrum of the measured complex optical field of a sample illuminated from a certain incident angle can be mapped onto the surface of a hemisphere called the Ewald sphere in the 3-D Fourier space, in which the centre position is translated by the incident angle of the illumination beam. Because we measured 300 complex optical fields from various incident angles, the 3-D Fourier space is filled with the mapped 2-D Fourier spectra, and taking the 3-D inverse Fourier transform of the filled 3-D Fourier space yields the 3-D RI distribution of the sample. Due to the finite NA of the detecting objective lens and condenser lens, the 3-D Fourier space has missing information called the missing cone, which is filled by an iterative non-negativity constraint algorithm[59]. The theoretical lateral and axial resolution of the system used in this work are 118.9 nm and 315.7 nm, respectively, which are determined from the maximum range of the reconstructed 3-D Fourier spectra[27]. The experimentally measured lateral and axial resolution are 270 nm and 920 nm, respectively, which were calculated by analysing the edge of the reconstructed tomograms of polystyrene beads. The visualization of 3D maps were performed using a commercial software (TomoStuido, Tomocube, Inc., Republic of Korea).

**Extraction of quantitative properties.** The reconstructed 3-D RI distribution of individual cells provides quantitative biochemical and structural information of cells including volume, protein concentration, and protein mass (dry mass). In order to measure biological properties of the cytoplasm and LDs separately, cell regions and LDs are segmented by the RI threshold of $n$ > 1.34 and $n$ > 1.375, respectively. The volume of the LDs and cytoplasm inside hepatocytes is calculated by integrating all voxels in the selected regions. In biological samples, the RI value is linearly proportional to the protein concentration inside cells as $n(x,y,z)=n_m+ \alpha C(x,y,z)$[20,21], where $n(x,y,z)$ is the 3-D RI distribution of samples, $n_m$ is the RI value of the surrounding medium ($n_m$ = 1.337 at $\lambda$ = 532 nm), $\alpha$ is an RI increment ($\alpha$ = 0.190 mL/g for protein and $\alpha$ = 0.135 mL/g for lipid[60]), and $C(x,y,z)$ is the protein concentration of the cellular regions or the lipid concentration of individual LDs. Thus, the protein concentration of the cytoplasm and the lipid concentration of LDs can be calculated with the measured RI contrast. Moreover, the dry mass, $m$, of LDs and cells is calculated by integrating calculated concentration as follows:

$$m = \iiint C(x,y,z)\,dx\,dy\,dz = \sum C(x,y,z)\Delta x \Delta y \Delta z = \frac{1}{\alpha}\sum \left[ n(x,y,z) - n_m \right]\Delta x \Delta y \Delta z \quad \text{(Eq.1)}$$

where $\Delta x$, $\Delta y$, and $\Delta z$ are the pixel resolution along the $x$, $y$, and $z$-axis, respectively.

**Acknowledgements**
The authors acknowledge Prof. Jennifer H. Shin and Dr. Mina Kim of the Korean Advanced Institute of Science and Technology for fruitful discussions. This work was supported by KAIST, KAIST Institute for Health Science and Technology, National Research Foundation (NRF) of Korea (2015R1A3A2066550, 2014K1A3A1A09063027, 2012-M3C1A1-048860, 2014M3C1A3052537), Tomocube Inc., and the Innopolis foundation (A2015DD126). K. K. acknowledges support from Global Ph.D. fellowship from NRF.

**Author contributions**
K. K., S. L and J. H. performed experiments. K. K. analysed the data. J. Y. developed the hepatocyte model. Y. P. and C. C. conceived and supervised the project. All the authors wrote the manuscript.

**Competing Financial Interests**
Prof. Park has financial interest in Tomocube Inc., a company that commercializes optical diffraction tomography instruments and phase imaging instruments and is one of the sponsors of the work.

**Figures and Captions**

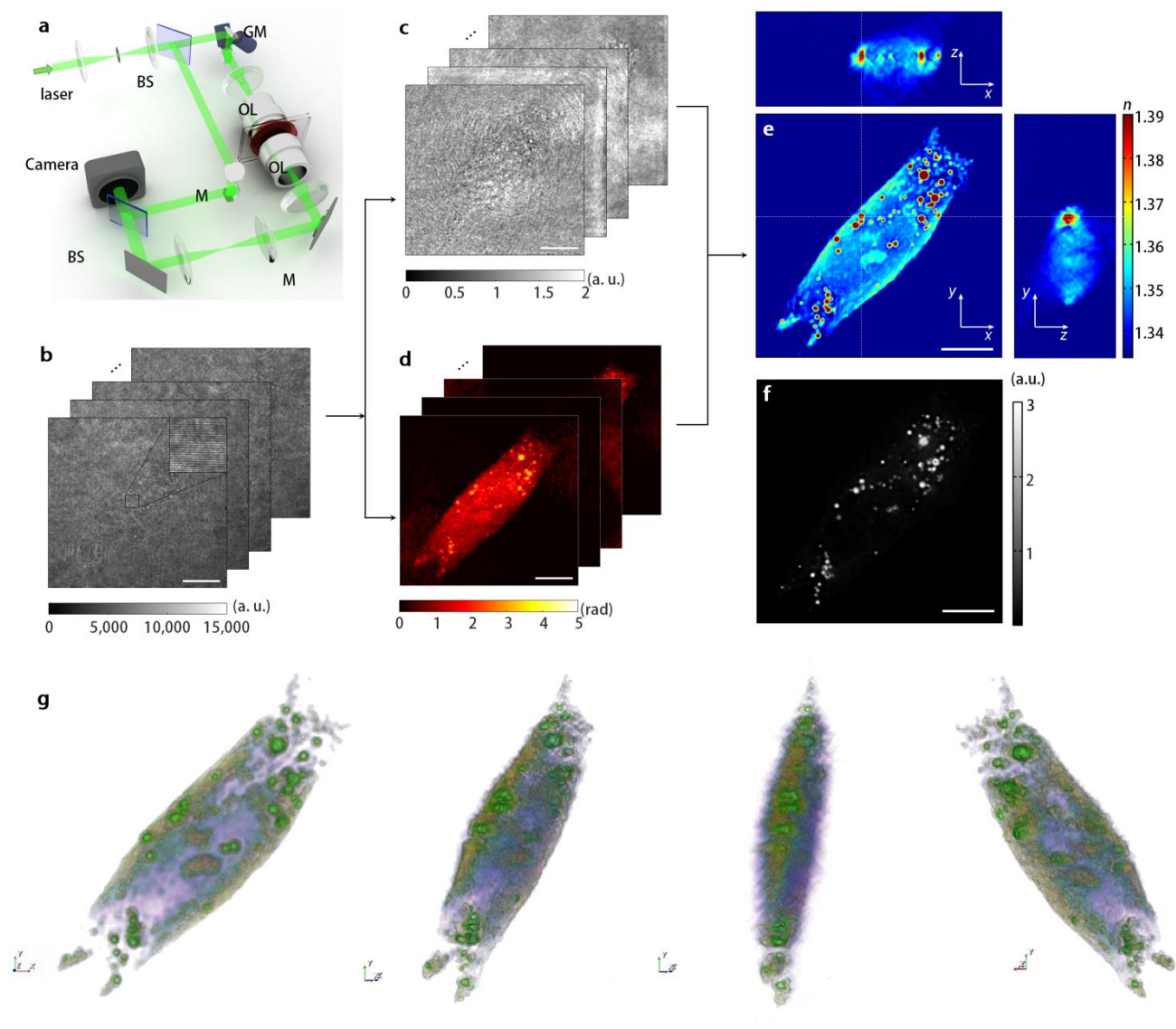

**Figure 1 | Schematic diagram of optical diffraction tomography optical diffraction tomography for label-free 3-D imaging of LDs. a**, Experimental set up. BS: beam splitter, GM: galvanomirror, M: mirror, OL: objective lens. **b**, Measured holograms of a hepatocyte (Huh-7) with various incident angles of illumination. The inset shows the enlarged view of the hologram indicated as a black box, which exhibits spatially modulated fringes. **c-d**, Retrieved amplitudes (**c**) and phases (**d**) of the sample illuminated from various incident angles. **e**, The cross-sectional slices of the 3-D RI distribution of the sample along the *x-y*, *y-z*, and *x-z* planes. **f**, A deconvoluted fluorescence image of LDs stained with Nile Red. The scale bars indicate 10 μm. **g**, 3-D rendered isosurface image of 3-D RI distribution from various viewing angles. See also Supplementary Movie 1.

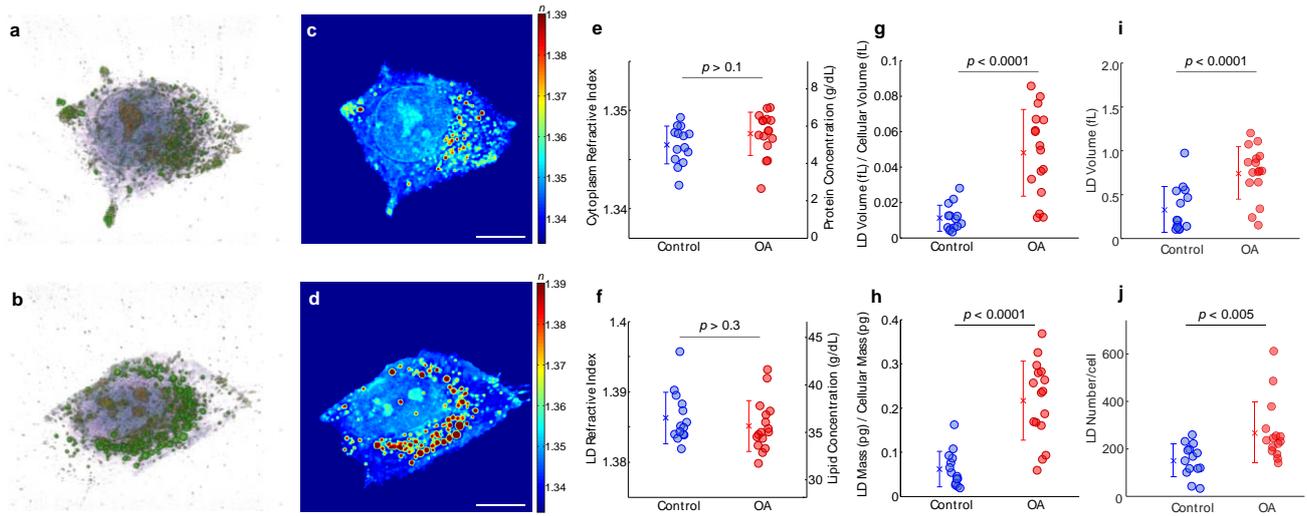

**Figure 2 | Quantitative analysis for LD formation in hepatocytes under oleic acid treatment. a-b,** 3-D rendered isosurface image of 3-D RI distribution of (**a**) an untreated and (**b**) OA treated hepatocyte. See also Supplementary Movie 2 and 3. **c-d**, Cross-sectional slice images of 3-D RI distribution of (**a**) the untreated and (**b**) OA treated hepatocyte. The scale bars indicate 10 μm. **e-j**, Quantitative analysis of (**e**) RI of cytoplasm, (**f**) RI of LDs, (**g**) the ratio of LD volume to cell volume, (**h**) the ratio of LD mass to cellular dry mass, (**i**) volume of individual LDs, and (**j**) number of LDs in untreated (*n* = 14) and OA-treated hepatocytes (*n* = 16).

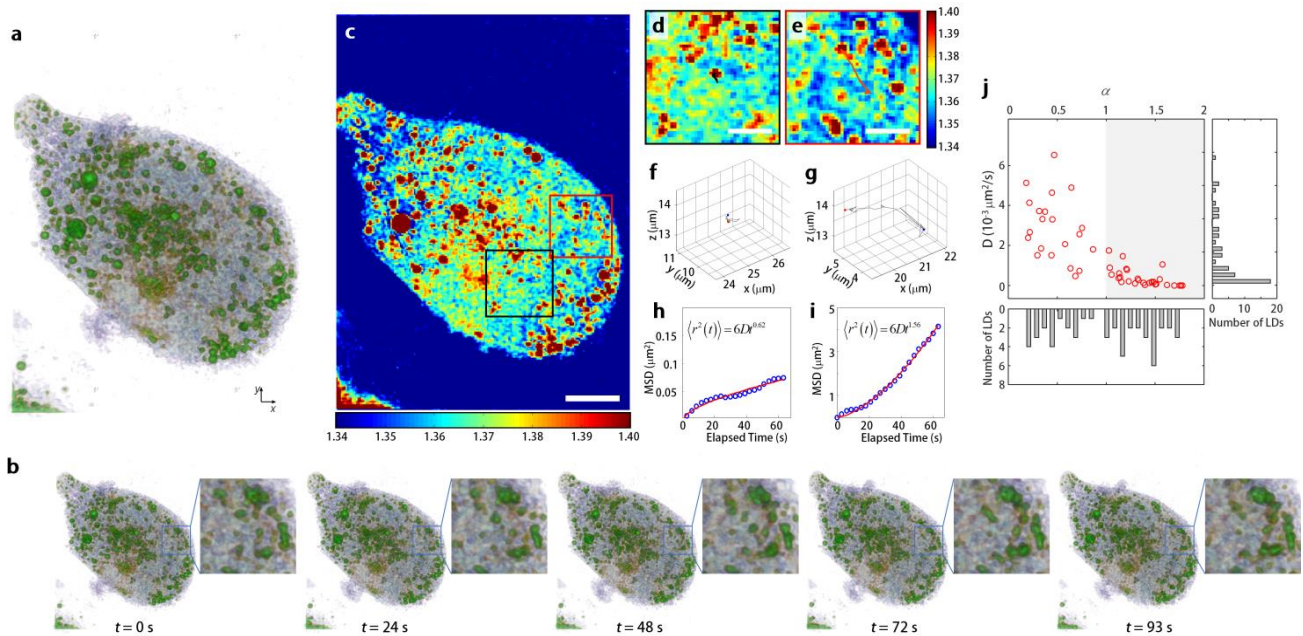

**Figure 3 | 4-D dynamics of individual LDs in hepatocytes. a-b**, The time-lapse 3-D rendered isosurface images of 3-D RI distribution of a hepatocyte. Insets in **b** show enlarged isosurface images of LDs indicated in a red box in **c**. See also Supplementary Movie 4. **c**, The cross-sectional slice image of the 3-D RI distribution of the hepatocyte along the *x-y* plane. Scale bar indicates 5 μm. **d-e**, Enlarged images of RI distribution of individual LDs indicated in a black (**d**) and red box (**e**) in **c**, respectively. The trajectories of the motion of LDs in the *x-y* plane for 93 seconds are indicated as black and red lines. Scale bar indicates 1 μm. **f-g**, The 3-D trajectories of LDs marked in **d** and **e**, respectively. **h-i**, Mean squared displacement along time of LDs in **d** and **e**. Experimental results are indicated as blue circles, and red lines exhibit the power-law fitted curves. **j**, Diffusion coefficients and power exponents of the power-law fitted curves taken from 51 individual LDs in 5 hepatocytes.